\documentclass{aa}

\usepackage{amsmath}
\usepackage{amssymb}
\usepackage{xcolor,subcaption}
\usepackage[export]{adjustbox}
\usepackage{flushend}
\usepackage[utf8]{inputenc}

\usepackage[colorlinks=true,linkcolor=blue,citecolor=blue,filecolor=blue,urlcolor=blue]{hyperref}

\usepackage[compact]{titlesec}  
\titlespacing{\section}{0pt}{12pt}{7pt}
\titlespacing{\subsection}{0pt}{9pt}{4pt}

\newcommand{\frb}{FRB~20180916B}

\usepackage[T1]{fontenc}

\DeclareRobustCommand{\VAN}[3]{#2}
\let\VANthebibliography\thebibliography
\def\thebibliography{\DeclareRobustCommand{\VAN}[3]{##3}\VANthebibliography}

\usepackage{graphicx}	% Including figure files
\usepackage{subcaption} % for them sub-floats
\usepackage{booktabs}   % beautiful tables
\usepackage{amsmath}	% Advanced maths commands
\usepackage{amssymb}	% Extra maths symbols
\usepackage{newtxtext,newtxmath}
\usepackage{multirow}
\newcommand{\cmd}[1]{\textsc{#1}}
\newcommand{\g}{\textsc{gain}}
\newcommand{\dg}{\textsc{dgain}}
\newcommand{\dphase}{\textsc{dphase}}

\newcommand{\mdp}{$\phi$}

\defcitealias{21DongziPA}{DL21}
\defcitealias{23McKinvenR3}{RM23}
\defcitealias{24GopinathR3}{AG23}
\graphicspath{{figures/}}
\graphicspath{{pbursts/}}
\title{Rotation Measure study of \frb\ with the uGMRT}
\titlerunning{\frb~RM with uGMRT}

\author{
S. Bethapudi\inst{\ref{i:bonn}}\thanks{sbethapudi@mpifr-bonn.mpg.de} \and L. G. Spitler\inst{\ref{i:bonn}} \and D. Z. Li\inst{{\ref{i:princeton}}} \and V. R. Marthi\inst{\ref{i:pune}} \and M. Bause\inst{\ref{i:bonn}} \and R. A. Main\inst{\ref{i:bonn},\ref{i:mcgill},\ref{i:tsi}} \and R. S. Wharton\inst{\ref{i:jpl}}.
}
\institute{
Max-Planck-Institute for Radio Astronomy, Auf dem H{\"u}gel 69, Bonn, Germany, 53121\label{i:bonn} \and
Department of Astrophysical Sciences, Princeton University, Princeton, NJ 08544, USA\label{i:princeton} \and
National Centre for Radio Astrophysics, Ganeshkhind, Post bag~3, Pune, India, 411 007\label{i:pune} \and
Department of Physics, McGill University, 3600 rue University, Montr\'{e}al, QC H3A 2T8, Canada\label{i:mcgill} \and
Trottier Space Institute, McGill University, 3550 rue University, Montr\'{e}al, QC H3A 2A7, Canada\label{i:tsi} \and
Jet Propulsion Laboratory, California Institute of Technology, Pasadena, CA 91109, USA\label{i:jpl}
}

\abstract{Fast Radio Burst 20180916B is a repeating FRB whose activity window has a 16.34 day periodicity that also shifts and varies in duration with the observing frequency. 
Recently, \citet{23McKinvenR3}~reported the FRB has started to show secular Rotation Measure (RM) increasing trend after only showing stochastic variability around a constant value of -114.6 rad m$^{-2}$ since its discovery.
RM studies let us directly probe the magnetic field structure in the local environment of the FRB.
The trend of the variability can be used to constraint progenitor models of the FRB.
Hence, further study of the RM variability forms the basis of this work.
}%
{We aim to study the local environment of \frb. 
We do so by focusing on polarization properties, namely RM, and study how it varies with time. 
The data comes from the ongoing campaigns of \frb~using the upgraded Giant Metrewave Radio Telescope (uGMRT). 
The majority of the observations are in Band~4, which is centered at 650 MHz with 200 MHz bandwidth. 
Additionally, we have a few observations where we had simultaneous coverage in Band~4 and Band~5 (centered at 1100 MHz).
}%
{We apply a standard single pulse search pipeline to search for bursts. 
In total, we detect 116 bursts with  $\sim$36 hours of on-source time spanning 1200 days, with two bursts detected during simultaneous frequency coverage observations.
We develop and apply a polarization calibration strategy suited for our dataset.
On the calibrated bursts, we use QU-fitting to measure RM.
We verify the veracity of calibration solution and RM measurement by performing RM measurements on single pulses of PSR~J0139+5814.
Lastly, we also measure various other properties such as rate, linear polarization fraction and fluence distribution.
}%
{Of the 116 detected bursts, we could calibrate 79 of them. 
From which, we observed in our early observations the RM continued to follow linear trend as modeled by \citep{23McKinvenR3}. 
However, our later observations suggest the source switch from the linear trend to stochastic variations around a constant value of -58.75 rad m$^{-2}$.
It has ceased any secular variability and is only showing stochastic variability.
Using the predicted Milky Way RM contribution, we report a tentative detection of a sign flip in the RM in the host galaxy, host-frame.
We also study cumulative rate against fluence and note that rate at higher fluences ($> 1.2$ Jy ms) scales as $\gamma=-1.09(7)$ whereas that at lower fluences (between $0.2$ and $1.2$ Jy ms) only scales as $\gamma=-0.51(1)$, meaning rate at higher fluence regime is steeper than at lower fluence regime.
Lastly, we qualitatively assess the two extremely large bandwidth bursts that we detected in our simultaneous multi~band observations. 
}%
{Future measurements of RM variations would help place stronger constraints on the local environment. Moreover, any periodic behavior in the RM measurements would directly test progenitor models. Therefore, we motivate such endeavors.}

\date{Received date / Accepted date}

\begin{document}
%\linenumbers
%\label{firstpage}
\maketitle
%%%%%%%%%%%%%%%%%%%%%%%%%%%%%%%%%%%%%%%%%%%%%%%%%%
%%%%%%%%%%%%%%%%% BODY OF PAPER %%%%%%%%%%%%%%%%%%
\section{Introduction}
% open with FRBs
\par Fast Radio Bursts (FRBs) are short timescale ($\sim$ms) transient events at radio wavelengths \citep{lorimer2007,19PetroffReview,22PetroffReview,22BailesReview}.
Based on repeatability, there are two classes of FRBs: repeating FRBs (rFRBs) and apparent non-repeating FRBs (nrFRBs).
Detecting multiple bursts from the same source enables detailed and cumulative studies of the FRB.
For example, rFRBs have been localized post-discovery by various interferometers, including the European Very Large Baseline Inteferometry Network \citep[EVN]{chatterjee2017,20MarcoteR3,KirstenM81,NimmoR67}. 
For select rFRBs that showed scintillation, detecting multiple bursts at multiple epochs enabled constraining the scintillation screen location \citep{22MainSTS,22OckerRfast}, and even led to modeling the effect of orbital motion of Earth on the scintillation timescales \citep{23MainAnnual,24WuR67Scint}.
Moreover, \citet{24KumarScin} and \citet{24NimmoScint} show how scintillation can place constraints on the location of the FRB emission region.
Exhaustive follow up of rFRBs, even with small telescopes, enables detecting extremely rare bright events and can help in suggesting a link between rFRBs and nrFRBs on the basis of energy distributions \citep{24KirstenLink}.
Similar follow-ups with sensitive instruments facilitates performing large number statistics on bursts from various rFRBs
\citep{21LiBimodal,22JahnsR1,22HewittR1,22LanmanR67,22ZhouR67}.
The repeating nature makes it easy to perform a fine time resolution study of bursts, which can probe the underlying emission mechanism \citep{20DayASKAP,21NimmoR3,21NimmoM81,23HewittR117,23SneldersR1}.
%Lastly, not just rFRBs, but localized nrFRBs also drive science by probing baryon matter between galaxies \citep{macquart2018} by placing constraints on the Hubble constant \citep{22JamesH0,22SteffenH0}.
Certainly, rFRBs allow us to study the FRB phenomena in a much greater detail, which would help us understand the FRB phenomena.
This paper focuses on a singular rFRB source and presents polarimetric results.

% what is Faraday Rotation? what is Rotation Measure?
%% maybe this is sticking out like a sore thumb
%%% idk how to put this in flow
%%%% cite the Faraday paper in thesis
%\par The rotation of the plane of polarization of the light which is travelling through a medium with magnetic field parallel to the light itself was first discovered by Faraday \citep[2152]{faradayrot}. \SB{this line is too long}
%The amount of rotation depends on the wavelength of light and a parameter known as Rotation Measure (RM). RM is a measure of free electron density and the magnetic field along the line of sight. 
%RM studies are one robust way to understand the magnetic fields between the observer and the sources of interest. Magnetic fields along the line of sight provide insights into the various characteristics of the medium.
%Hence, it becomes particularly important in studying the Rotation Measures of Fast Radio Bursts (FRBs) as an attempt to study magnetic fields at all \weird{places}.
%FRBs are bright, extragalactic transients detected. 

%% 

% recent magnetoionic-dynamic environments
% first for R3
\par \frb~is a rFRB discovered by the Canadian Hydrogen Intensity Mapping Experiment/Fast Radio Burst (CHIME/FRB) experiment \citep{chime2019b}.
It is unique in that its activity windows are periodic with a $16.34$ day period \citep{20R3Period}, and shift and vary in length with observing frequency \citep{21ZiggyR3,21PastorR3,23BethapudiR3}.
In other words, the source exhibits periodicity and chromaticity.
%% describe the host galaxy Tendulkar+ and Kaur+ results
The source was localized to an edge of a spiral galaxy at a redshift of $0.0337$ with EVN \citep{20MarcoteR3}. \citet{21TendulkarR3}, using Hubble Space Telescope (HST) observations of the host galaxy, placed constraints on the age of the FRB source to be of the order of 10 Myr.
\citet{22KaurR3}~performed HI spectroscopy observations using the upgraded Giant Metrewave Radio Telescope (uGMRT) with which they report the host galaxy to be gas-rich and have a low Star Formation Rate (SFR), suggesting the host galaxy might have undergone a minor merger in recent past.
%\SB{what is this?}The close proximity of the FRB source to the distorted velocity field which is caused by minor merger could not just be a coincidence \citep{22KaurR3}. 
Future sensitive observations would further illuminate the local environment and provide an additional constraint on possible source models.

%%% RM results of R3
\par \citet[hereafter \citetalias{23McKinvenR3}]{23McKinvenR3}, using CHIME/FRB, report that Rotation Measure (RM) of bursts from \frb~has started to change. 
The RMs are consistent with a single value from its discovery until around MJD~59243, but since then, they have started to exhibit an increasing trend.
The trend is further corroborated by LOFAR detections \citep[hereafter \citetalias{24GopinathR3}]{24GopinathR3}.
\frb~is not the only rFRB to exhibit RM variability.
\citet{23McKinvenFRB}~report nearly all repeaters exhibit RM variability, which suggests repeaters exist in dynamic magneto-ionic environments.
This variability which is local to the repeater can be used to constrain the origins of FRB.
See, for example \citet{23YangVRM}, which predicts RM variability for different FRB progenitor scenarios. 
However, \citet{23ShermanHG}~suggest that host galaxies of rFRBs have statistically larger magnetic fields than that of the Milky Way, that is, suggesting host galaxy ISM to contribute significantly to the FRB RM.
In such a case, decoupling host galaxy contribution and local environment would be particularly challenging, but nevertheless will be crucial to place limits on local environments.

\par This paper mainly presents polarization results from the ongoing observing campaign of \frb~with the upgraded Giant Metrewave Radio Telescope (uGMRT). 
Sect.~\ref{sec:obssearch}~describes the observations and the search strategy utilized. 
%Then, Sec.~\ref{sec:cal} covers the polarization calibration (Sec.~\ref{ssec:polcal}), flux scaling (Sec.~\ref{ssec:fluxcal}), PA extraction (Sec.~\ref{ssec:pacorr}), and RM measurement in Sec.~\ref{ssec:rmm}.
Then, Sect.~\ref{sec:analysis} covers the polarization calibration, flux scaling and RM measurements.
Lastly, results and discussions are contained in Sect.~\ref{sec:rd}, with conclusions in Sect.~\ref{sec:con}.
%Results pertaining with Position Angle will be presented in a separate accompanying paper.

\section{Observations and searches} \label{sec:obssearch}

\subsection{Observations}

\par Observations were performed using uGMRT \citep{gmrt1991,ugmrt}. 
It is a radio interferometer situated in Khodad, near Pune, India. 
It consists of thirty 45-meter parabolic dishes which can be used to perform imaging- and phased array-mode observations. It offers observation capability from $100$ MHz to $1.4$ GHz broken into five bands. 
Our observations used Band~4 (550 to 750 MHz) and Band~5 (1000 to 1200 MHz).
% cycle-39, 42, 43
Observations from GMRT Proposal Cycle-39 (from October-2020 to March-2021), Cycle-43 (from October-2022 to April-2023) and Cycle-45 (from October-2023 to April-2024), as well as one observation from Cycle-42 (from April-2022 to October-2022), are presented in this paper.
Each cycle has its own observing strategy, chiefly the time resolution and frequency coverage. 
All observations are primarily done in Band~4.
However, for five observations of Cycle-39, the entire GMRT array was split into two sub-arrays, where one sub-array performed observations in Band~4 and the other in Band~5.
This way, simultaneous coverage of Bands~4 and~5 was achieved.
%However, five observations of Cycle-39 had both Band~4 and Band~5 coverage, which is made possible by the sub-arraying capability of uGMRT.
All observations of Cycles-39, -43, and -45 have time resolution of $327.68~\mu$s, whereas, Cycle-42 observations only have time resolution of $168.34~\mu$s.
The number of channels in all the observations was fixed to 2048.
We note that all observations of Cycle-45 have been strongly affected by strong ionospheric activity. 
%% this para bad
%All observations included quasar (3C138) scans and two test pulsars - B0329+54 and J0139+5814 scans. However, early observations from Cycle-39 only consisted of 3C138 quasar scans. Sometimes, we did not have any quasar/pulsar scan. 

\begin{table}
\centering
\caption{Band~4 Observations described in this paper. Date and Time are specified in UTC. $T_{\rm obs}$ is duration in hours. $N$ is the number of bursts. Aux. column presents the auxiliary sources observed. Dash represents no source and NG denotes noise diode scan.
We only show few select observations. Please see auxiliary material for machine-readable csv file.}
%%% MANUAL ADJUSTMENT
\resizebox{\columnwidth}{!}{%
\begin{tabular}{@{}ccccc@{}} 
\toprule \toprule
 Date & Time  & $T_{\rm obs}$& $N$ & Aux.\\ \midrule
%2020-11-06  &  19:58:52.724  &  0.92 & - & -\\
%  &  21:06:24.758  &  0.59 & 1 & -\\
%2020-11-08  &  21:43:59.952  &  0.92 & 2 & -\\
%  &  22:44:44.634  &  0.92 & 2 & -\\
%2020-11-09  &  20:20:31.789  &  0.91 & - & -\\
%  &  21:34:30.369  &  0.93 & - & -\\
%2020-12-08 &  19:38:11.003  &  0.92 & 4 & \multirow{2}{*}{3C138}\\
%  &  20:38:44.277  &  0.76 & - & \\
%2020-12-10  &  20:59:51.346  &  0.51 & - & -\\
%2020-12-11  &  19:50:48.112  &  0.85 & - & -\\
%  &  20:45:18.327  &  0.76 & - & -\\
%% where is 11jan2021?
%2021-01-28  &  14:36:45.729  &  0.92 & 2 & \multirow{2}{*}{B0329+54}\\
%  &  15:43:53.603  &  0.53 & 2 &  \\
2021-01-29 &  09:41:44.132  &  0.92 & 7 & \multirow{4}{*}{3C138, B0329+54}\\
  &  10:47:48.924  &  0.92 & 16 &  \\
  &  11:52:26.474  &  0.92 & 5 &  \\
  &  12:57:30.196  &  0.92 & 13 &  \\
%2021-01-30  &  14:19:46.324  &  0.92 & 5 & \multirow{2}{*}{3C138, B0329+54}\\
%  &  15:23:42.938  &  0.92 & 1 &  \\
%2021-03-01  &  13:38:24.309  &  0.92 & 1 & \multirow{2}{*}{\shortstack[l]{3C138, B0329+54,\\ J0139+5814}}\\
%  &  14:46:00.369  &  0.26 & - &  \\
%2021-03-02  &  11:15:14.014  &  0.75 & 3 & \multirow{2}{*}{\shortstack[l]{3C138, B0329+54,\\ J0139+5814}}\\
%  &  12:08:42.489  &  0.75 & 4 & \\
%2021-03-04  &  13:19:59.534  &  0.50 & - & -\\
%  &  13:58:30.763  &  0.50 & - & -\\
%  &  14:36:59.979  &  0.50 & - & -\\
%  &  15:15:29.866  &  0.50 & - & -\\
%2021-03-16  &  08:33:52.918  &  0.67 & - & \multirow{2}{*}{\shortstack[l]{3C138, NG,\\ B0329+54}}\\
%  &  09:25:44.085  &  0.66 & - &  \\
%2021-03-19  &  07:19:34.109  &  0.75 & 2 & -\\
%  &  08:12:22.318  &  0.75 & 1 & -\\
%  &  09:06:48.507  &  0.47 & 2 & -\\
%2021-12-20  &  16:54:34.688  &  0.50 & - & \multirow{3}{*}{3C138, B0329+54}\\
%  &  17:40:45.613  &  0.56 & 7 &  \\
%  &  18:22:46.893  &  0.51 & 3 &  \\ 
2022-11-11  &  18:58:05.567  &  0.37 & 2 & \multirow{6}{*}{\shortstack[l]{3C48, 3C138,\\ NG, B0329+54,\\ J0139+5814}}\\
  &  19:33:08.088  &  0.37 & - &  \\
  &  20:08:13.293  &  0.37 & 2 &  \\
  &  20:43:39.302  &  0.37 & - &  \\
  &  21:18:47.191  &  0.37 & - &  \\
  &  21:54:27.293  &  0.37 & - &  \\
%2022-12-31 &  13:34:13.883  &  0.37 & 1 & 3C48, 3C138, NG, B0329+54, J0139+5814 \\
%2022-12-31 &  13:34:13.883  &  0.37 & 1 & \multirow{6}{*}{\shortstack[l]{3C48, 3C138,\\ NG,B0329+54,\\ J0139+5814}} \\
%  &  14:12:07.532  &  0.37 & 1 &  \\
%  &  14:49:16.888  &  0.37 & - &  \\
%  &  15:27:14.563  &  0.37 & 1 &  \\
%  &  16:04:20.564  &  0.37 & 1 &  \\
%  &  16:41:30.592  &  0.37 & - &  \\
%2023-02-18 &  10:48:53.517  &  0.37 & 1 & \multirow{6}{*}{\shortstack[l]{3C48, 3C138,\\B0329+54,\\J0139+5814}} \\
%  &  11:29:02.725  &  0.37 & 2 &  \\
%  &  12:09:16.631  &  0.37 & 1 &  \\
%  &  12:49:26.510  &  0.37 & 1 &  \\
%  &  13:29:40.416  &  0.37 & 1 &  \\
%  &  14:09:52.979  &  0.10 & 1 &  \\ 
2023-12-25  &  09:20:42.255  &  0.37 & 4 & \multirow{4}{*}{\shortstack[l]{3C468.1,NG,\\J0139+5814}}\\
  &  09:58:07.717  &  0.37 & 3 \\
  &  10:43:28.982  &  0.37 & 4 \\
  &  11:23:30.137  &  0.37 & 6 \\
  &  12:03:31.963  &  0.10 & - \\ \bottomrule
%2024-01-27  &  07:24:00.898  &  0.37 & - & \multirow{4}{*}{\shortstack[l]{3C468.1,NG,\\J0139+5814}}\\
%  &  08:03:03.668  &  0.37 & - \\
%  &  08:41:42.950  &  0.37 & - \\
%  &  09:20:20.219  &  0.37 & - \\
%2024-02-12  &  06:19:21.171  &  0.33 & - & \multirow{4}{*}{\shortstack[l]{3C468.1,NG,\\J0139+5814}}\\
%  &  06:56:32.540  &  0.33 & - \\
%  &  07:32:58.947  &  0.33 & - \\
%  &  08:09:29.380  &  0.33 & 1 \\
%  &  08:46:10.551  &  0.33 & - \\ \bottomrule
\end{tabular}%
}
\label{tab:band4}
\end{table}

\begin{table}
\centering
\caption{Band~5 Observations described in this paper. Date and Time are specified in UTC. $T_{\rm obs}$ is duration in hours. $N$ is the number of bursts.}
\resizebox{\columnwidth}{!}{%
\begin{tabular}{@{}cccc@{}}
\toprule \toprule
 Date & Time  & $T_{\rm obs}$& $N$ \\ \midrule
2020-11-06  &  21:06:24.758  &  0.59 & 1 \\
2020-12-08  &  19:38:12.345  &  0.92 & - \\
  &  20:38:44.277  &  0.76 & 2 \\
2021-01-28  &  14:36:45.729  &  0.92 & 2 \\
  &  15:43:53.603  &  0.53 & 1 \\
2021-03-01  &  13:38:23.638  &  0.92 & 1 \\
  &  14:46:00.369  &  0.26 & - \\
2021-03-16  &  08:33:52.918  &  0.67 & - \\
  &  09:25:44.085  &  0.66 & - \\ \bottomrule
\end{tabular}%
}
\label{tab:band5}
\end{table}

\par Each individual scan is listed in Table~\ref{tab:band4} for Band~4 and Table~\ref{tab:band5} for Band~5. 
In all the cycles, we recorded full coherency 16-bit Phased Array (\cmd{pa}) filterbank and Stokes-I 16-bit Intensity Array (\cmd{ia}) filterbank \citep{gmrtbeamform}.
\cmd{pa}-beam is generated using voltages from individual antennas which are summed in phase, thus is a \emph{coherent} beam. 
Whereas, \cmd{ia}-beam is constructed using Total Intensities of each antenna.
All the observations are scheduled in the active windows as predicted by the periodicity and chromaticity model \citep{20R3Period,21ZiggyR3,21PastorR3,23BethapudiR3}.

\par Each observation consists of science scans, which are bracketed by phasing scans of 0217+738. 
%These phasing scans are essential for using uGMRT, which is an interferometer, as a phased array telescope. For all the observations, we used 0217+738 as the phase calibrator. 
Unfortunately, our choice of calibrators and pulsars (for verifying calibration pipeline) has not been consistent throughout the cycles. 
All together, we observed 3C138 (a polarized quasar), 3C48 and 3C468.1 (unpolarized quasars), noise diode scans (NG), and pulsars: PSR~B0329+54 and PSR~J0139+5814.
Additionally, for some of the observations, we have noise diode scans during which we pointed at an empty patch of sky with noise diode turned ON. 
We specify the auxiliary sources in the Aux. column of Table~\ref{tab:band4}.
When we discuss various calibration strategies in Sect.~\ref{ssec:polcal}, we use whichever strategy is possible for an observation based on the available source scans.

%%%%%%%%%%%%%%%%%%%%%%
%% table should contain
%% epoch of observation, science-observation-span, number of bursts detected, quasar scan, pulsar scans, noise diode scan
%%%% do i give individual start times of each of the scans?
%%%%%
%%%%%%%%%%%%%%%%%%%%%%

\subsection{Searches}

\par We subtract the time aligned \cmd{ia}-beam from the \cmd{pa}-beam and create a Stokes-I 8-bit filterbank data. 
Doing so gets rid of all the RFI common in both beams, while leaving the astrophysical signal, which is only dominantly present in the \cmd{pa}-beam, untouched. We then apply a standard \cmd{presto} \citep{presto}~single pulse search pipeline. 
Post subtraction, the data is already clean that we do not need to apply any additional Radio Frequency Interference (RFI) mitigation step \citep{gmrtbeamform}. 
We search at DM of $348.82$ pc cm$^{-3}$ at the native time resolution. We search for candidates which are up to 300 samples wide ($100$ ms). 
We manually vet each of the candidates using a diagnostic plot.
For each of the manually vetted candidates, we slice the original 16bit \cmd{pa}-beam data to create a \cmd{psrfits} burst archive \citep{psrchive}. 
%These burst archives are populated with correct uGMRT system characteristic, such as feed basis, frequency axis, polarization type, and so on. 
All the analysis from now on is carried out on these burst archives. The number of bursts we report are listed in Table.~\ref{tab:band4} for Band~4 and Table~\ref{tab:band5} for Band~5.

\par We report a total of $116$ bursts in Band~4 and seven bursts in Band~5. 
Out of the $116$ Band~4 bursts, we were only able to polarization calibrate $79$ of them, because of lack of auxiliary sources in some observations (see Tab.~\ref{tab:band4}).
Moreover, we also report two simultaneous detections of bursts in Bands~4 and~5. 
Unfortunately, the Band~4 bursts of the two simultaneous detections could not be calibrated.
We defer polarization calibration of Band~5 bursts to future work.

\section{Analysis} \label{sec:analysis}

%%%%%%%%%%%%%%%%%%%%%%%%%%%%%%%%%%%%
\begin{table*}
%\begin{longtable}
    \centering
    \caption{Properties of the Band~4 bursts detected in this work. MJD refers to topocentric UTC time of the burst when de-dispersed at $348.820$ pc cm$^{-3}$ with 750 MHz as the reference frequency. Bandwidth is the frequency envelope of the bursts, measured in MHz. RM refers to Rotation Measure of the bursts. Linear refers to linear polarization fraction in percentage. Note that it could be more than $100\%$ because we have removed the baseline using \cmd{psrchive}. 
    Width is the time span of the ON-region measured in ms. 
    Flux, measured in mJy, refers to the peak flux in the frequency-averaged flux-calibrated time series.
    Fluence is the sum over the ON-region of the frequency-averaged flux-calibrated time series, which is measured in Jy ms.
    RM$_{\rm ion}$ denotes the ionospheric RM contribution seen by uGMRT at the epoch of burst.
    For brevity, we only display the first ten entries here. Please find the rest in auxiliary material.
    }
\resizebox{\textwidth}{!}{%
    \begin{tabular}{cccccccccccc}
    \toprule \toprule
    Burst & MJD & Width & Bandwidth & Fluence & Flux & RM & RM error & Linear & RM$_{\rm ion}$ \\
    & & ms & MHz & Jy ms & mJy & rad m$^{-2}$ & rad m$^{-2}$ & \% & rad m$^{-2}$ \\
    \midrule
    %\endfirsthead
    %\caption{continued.}\\
    %Burst & MJD & Width & Bandwidth & Fluence & Flux & RM & RM error & Linear & PA & PA error & RM$_{\rm ion}$ \\
    %& & ms & MHz & Jy ms & mJy & rad m$^{-2}$ & rad m$^{-2}$ & \% & deg & deg & rad m$^{-2}$ \\
    %\midrule \midrule
    %\endhead
    %% reduce the precision of width and bandwidth
B1 & 59191.828612 & 12.8 & 122.7 & 0.40 & 0.96 & -121.12 & 1.74 & 84.2 & -0.2 \\
B2 & 59191.834280 & 10.2 & 175.6 & 0.26 & 0.57 & -118.80 & 1.45 & 64.1 & -0.2 \\
B3 & 59191.840990 & 10.2 & 169.2 & 0.46 & 1.02 & -117.20 & 0.90 & 87.6 & -0.2 \\
B4 & 59191.851290 & 15.4 & 121.0 & 0.86 & 2.92 & -115.22 & 0.68 & 98.2 & -0.2 \\
B5 & 59243.408788 & 8.9 & 101.0 & 0.17 & 0.34 & -111.34 & 3.50 & 46.5 & -0.4 \\
B6 & 59243.413121 & 16.1 & 60.0 & 0.47 & 0.85 & -116.39 & 1.38 & 81.0 & -0.4 \\
B7 & 59243.415825 & 4.9 & 38.7 & 0.37 & 0.80 & - & - & - & -0.3 \\
B8 & 59243.427116 & 3.0 & 85.3 & 0.33 & 0.67 & -116.48 & 3.70 & 50.5 & -0.3 \\
B9 & 59243.431151 & 10.8 & 111.5 & 0.28 & 0.49 & -110.58 & 1.85 & 53.5 & -0.3 \\
B10 & 59243.436477 & 3.6 & 140.8 & 0.96 & 1.85 & -115.78 & 0.69 & 89.4 & -0.3 \\ \bottomrule
    \end{tabular}
}
    \label{tab:bursts}
\end{table*} 
%\end{longtable} 
\par Having created \cmd{psrfits} burst archive from the \cmd{pa}-beam data, we first employ \cmd{pazi} \citep{psrchive}~to perform extensive RFI cleaning. Then, we visually identify the ON-burst region as a rectangular patch in time and frequency. 
The ON-burst region in time is measured to be the width, and ON-region in frequency is measured to be the bandwidth of the burst.
%All analyses which uses ON-burst and OFF-burst region use the rectangle we identify here.
We then perform polarization calibration and flux scaling as described in Sects.~\ref{ssec:polcal} and~\ref{ssec:fluxcal}.
%Thereafter, we perform flux scaling as described in Sect.~\ref{ssec:fluxcal} to measure fluxes. 
%We do so because our polarization calibration model already involves a scalar gain term per frequency channel which only needs to be appropriately scaled to measure fluxes.
Then, we perform RM, polarization fraction measurements as described in Sect.~\ref{ssec:rmm}.

\subsection{Polarization calibration} \label{ssec:polcal} 

\par uGMRT in Band~4 has circular basis, which has two hands: Right Circular Polarized (RCP) and Left Circular Polarized (LCP).
%Each observation has different set of auxiliary sources to assist in calibration (see Table~\ref{tab:band4}). 
To polarization calibrate, we require a polarized source like 3C138 (polarized quasar) or noise diode (NG). 
3C48 is unpolarized, however, we can use 3C48 in conjunction with any of pulsars, PSR~B0329+54 or PSR~J0139+5814, to perform calibration.
We judiciously choose calibration strategy based on whichever auxiliary observations are available (see Table~\ref{tab:band4}).
%For the observations of \texttt{09dec2020}, \texttt{29jan2021}, \texttt{30jan2021}, \texttt{01mar2021}, \texttt{02mar2021}, \texttt{20dec2021} we use 3C138.
%We use noise diode to calibrate observations of \texttt{12nov2022}, \texttt{31dec2022}.
%We use 3C48 and PSR~J0139+5814 to calibrate the observation of \texttt{18feb2023}.
Unfortunately, we also have few observations where no auxiliary source had been observed. 
We have no way of performing polarization calibration and are forced to omit them from polarization analysis.
%We also Position Angle correction exclusively using the PSR~B0329+54 scans. In case of \texttt{12nov2022}, we use PSR~J0139+5814 to perform correction.

%% only now singleaxis
\par \cmd{SingleAxis} is the simplest model to characterize the polarization response of the system. 
This model involves a scalar gain term (\g), which is a multiplicative factor; \dg,~which accommodates for the differential gain between the two hands; and \dphase,~which measures the cross hand delay, that is, the delay between the RCP and LCP.
We model \dphase($\phi$)~as a linear function of observing frequency (in GHz, denoted by $f_{\rm GHz}$)  as $\phi = \psi_{\rm r} + \pi{\rm D}_{\rm ns}f_{\rm GHz}$. 
The $\psi_{\rm r}$ is a constant term which, in a circular basis, is half the position angle of the infalling radiation.
%This position angle is not strictly a meaningful quantity because at this point we have not performed any parallactic corrections, but instead is only a fit parameter.
${\rm D}_{\rm ns}$ is the cross hand delay in nanoseconds and is also known as cable delay in the literature.

\par We make an assumption when computing \g~and \dg~using quasar scans that quasars possess no circular polarization \citep{PerleyButler2014Pol}. And, while the quasar flux density is not flat over frequency \citep{PerleyButler2017}, we assume the quasar flux density to be same throughout the bandwidth of $200$ MHz. 
We make this assumption because throughout the work, we average over the frequency and quote the measurements at one frequency (which is, the central frequency of $650$ MHz).
Since a noise diode is designed to emit constant pure Stokes-U signal across the bandwidth, the assumptions we make before automatically hold true.
Then, for all the quasar and noise diode scans, we subtract 
OFF (when looking at an empty patch of sky or when the noise diode is turned OFF) from ON (when looking at the quasar or when the noise diode is turned ON) scans and average over time to produce Stokes products against frequency. 
Using this, we measure \g~and \dg~algebraically \citep[see Eq. 14]{Britton2000}\footnote{Please note of an error in \citet{Britton2000} as mentioned in \texttt{SingleAxis.C} of \texttt{psrchive} code}. 

%% dphase measurements
\par We use Bayesian Nested Sampling code \cmd{ultranest} \citep{ultranest} to fit for ${\rm D}_{\rm ns}$ and $\psi_{\rm r}$.
The position angle-like term is fixed to be in $[ \frac{-\pi}{2}, \frac{\pi}{2} ]$. The ${\rm D}_{\rm ns}$ is restricted to be in $[-400, 400]$ ns. 
Note that the sign of the delay is not a consequence for calibration or RM measurements. 
We model \begin{equation}
    Q + jU = L_p\ I_{\rm smooth}\ {\rm exp} \big[  1j \big( \psi_{\rm r} + \pi{\rm D}_{\rm ns}f_{\rm GHz} \big) \big],
\end{equation}
where $L_p$ is the linear polarized fraction, $I_{\rm smooth}$ is the Gaussian-smoothed Stokes-I profile over frequency.
We use a standard Gaussian likelihood between observed data (Stokes-Q and Stokes-U against frequency) and the model predicted Stokes-Q and -U. 
Additionally, we treat standard deviation of the OFF scan to be the respective errors in Stokes-Q and -U, and use them in the likelihood.
Measuring the cross hand delay with noise diode scans is straightforward.
However, measuring the same using either the 3C138 quasar or pulsars requires a preceding step. 

%% 3C138 dphase measurements
%\par 
%\citet{22BaghelGmrtPol}~uses 3C138 to polarization calibrate the visibility data at Band~4 with uGMRT.
%Therein, 3C138 polarization fraction (and position angle) are modeled as a function of frequency by extrapolating \citet{PerleyButler2014Pol}.
%Since we only report measurements at the central frequency, we do not model the frequency variations.
%We assume a singular position angle value across the entire bandwidth and moreover, we also let polarization fraction to be single fit parameter over the bandwidth. 
\par While 3C138 is generally accepted to have zero RM \citep{PerleyButler2014Pol}, \citet{Cotton1997}~suggests RM to be $-2.1$ rad m$^{-2}$. 
Since we are operating at lower frequencies (compared to L-band and above where VLA operates) where the rotation is more enhanced, we have to take this Faraday Rotation into account.
In addition, there is also the ionospheric RM contribution, which also inadvertently causes Faraday Rotation.
The Faraday Rotation affects the measurement of cross hand delay (\mdp) as both are degenerate.
Then, if we incorrectly compensate for cross hand delay and apply such a solution to the bursts, we end up with incorrect RM measurements.
Therefore, prior to fitting for \mdp, we de-rotate Stokes-Q and Stokes-U accordingly, that is, we compensate for the spurious RM along the line-of-sight.
%% Pulsar RM
\par We also measure cross hand delay using two pulsars - PSR~B0329+54 and PSR~J0139+5814. 
Since the pulsars have position angles varying with phase, we do not average over the entire ON-region, instead pick a single phase bin which has the highest intensity.
We make use of this slice to compute cross hand delay.
As done in case of 3C138, we also correct for literature value of rotation measure and ionospheric RM along the line-of-sight.
We estimate the ionospheric RM contribution using \citet{rmextract}.

\subsection{Flux calibration} \label{ssec:fluxcal}

\par Flux calibration is performed using the quasar scans. \citet{PerleyButler2017} is used to estimate the flux scaling which is then applied to the bursts. Note that flux scaling does not depend on the polarization dimensions, and only affects the scalar \g~term. 
Therefore, we can perform flux scaling even for those observations for which we could not perform polarization calibration.

\par When calibration is done using quasars (either polarized or unpolarized), flux calibration is straightforward. 
When calibration is done using noise-diode scans, the \g~term is determined by the noise-diode whose strength is unknown. 
To circumvent this, we first apply the noise-diode derived calibration solution to 3C48 quasar scan and then determine the flux scaling from the calibrated 3C48 quasar scan.  
To those bursts that could not polarization calibrated, flux scaling can still be performed because only a known flux source is needed. 
After performing flux scaling to all the bursts, we measure the flux and fluence.
We first collapse the frequency axis by averaging over it.
Then, we sum over the ON-region to compute the fluence in Jy ms.
The maximum in the ON-region is measured as the peak flux in mJy.
Note that after flux calibration, the OFF-region statistics should be zero mean already, therefore, we do not have to do any OFF-region subtraction here.
We tabulate the measurements in Table~\ref{tab:bursts}.
We assume a 10\% error in our flux and fluence measurements as per \citet{ugmrtfluxerror}. For the observations of MJD~60303, MJD~60336 and MJD~60352 (GMRT cycle~45), we could not perform flux calibration because of the heightened ionospheric activity adversely affecting our observations.

% do i estimate noise diode strength? or naah??
% this is observatory work

\subsection{RM and polarization fraction measurements} 
\label{ssec:rmm}

%One way to measure RM is by observing that RM and $\lambda^2$ form a Fourier transform pair. This methodology is called RM synthesis \citep{05Brentjens}. 
% two ways to measure RM
\par Faraday Rotation introduces sinusoidal variations in Stokes-Q,-U as a function of $\lambda^2$, where $\lambda$ is the wavelength. 
We employ the method of $QU$-fitting to measure RM where we directly 
fit sinusoidals to Stokes-Q,-U as a function of $\lambda^2$. 
% Introduce QU-fitting
We model Stokes-Q,-U as a function of wavelength as
\begin{equation}
    \label{eq:rm}
    Q + jU = L_p\ I_{\rm smooth}\ {\rm exp} \big[ 2j \big( {\rm RM} (\lambda^2 - \lambda_c^2) + \psi \big) \big].
\end{equation} $L_p$ is the linear polarization fraction. 
$\psi$~is the Position Angle. 
$I_{\rm smooth}$ is the Gaussian smoothened Stokes-I profile as a function of frequency. $\lambda_c$ is the wavelength at the center of the burst frequency envelope. We perform the centering to assist in convergence of the fit. Without doing so, we noted a correlation between RM and $\psi$ in our posterior sampling. 

\par We sample from the posterior using Nested Sampling code \cmd{ultranest}~\citep{ultranest}. 
We use a standard Gaussian likelihood function with errors in Stokes-Q and -U, and assume flat priors for all the parameters. 
Our prior on RM ranges from $-400$ rad m$^{-2}$ to $400$ rad m$^{-2}$. 
And on $L_p$ from $0$ to $1$. 
The prior on $\psi$ is in $[-\frac{\pi}{2},\frac{\pi}{2}]$.
Nested Sampling algorithm samples from the posterior distribution directly. We take the median and standard-deviation of each of the parameters as the measured value and corresponding error. 
% describe the data analysis
For every burst, we time-average the ON-burst and OFF-burst regions in all Stokes channels. 
Then, we subtract OFF-burst from ON-burst and use ON-OFF Stokes-parameters (which are a function of wavelength) for further analysis.
We estimate the errors as the standard deviation of the OFF-burst regions.
We notice that the errors might be underestimated this way and furthermore any error in calibration can also introduce systematic in the error which otherwise would be left unaccounted for. Therefore, we introduce \cmd{equad} to the QU-fitting to artificially inflate the errors of Stokes-Q, -U. \cmd{equad} is an additional parameter that is fitted for while measuring RM. 
% the ranges
The prior of \cmd{equad} is also flat and is set to be in $[0.,10.]$. 
In case the posterior samples of \cmd{equad}~were converging to the boundary values, we re-ran the fitting after suitably increasing the \cmd{equad} range. 
For each fitting, we produced and visually inspected diagnostic plots. 

%% this after measurement
%% okaay
\par The measured RMs and corresponding error are tabulated in Table~\ref{tab:bursts}.
Lastly, we verify our RM measurements in the Appendix~\ref{a:sec:rmver}, and we comment about the \dphase~variability and its affect on RM measurement in Appendix~\ref{a:sec:dpvar}.

\subsubsection{Band~5}

% okay gmrt band-5
\par uGMRT Band~5 data is in linear basis. We do not perform any calibration in Band~5 because of lack of good quality calibrator data. Moreover, since we only have two bright bursts in Band~5, we leave the calibration work to future work. Fortunately, because of linear feeds, it is still possible to measure RM without calibration. 

% describe how
%\par Under \cmd{SingleAxis} model for linear basis, the boost and rotation axes is Stokes-Q. So, \dg~controls boost (hyperbolic rotation) between Stokes-I and Stokes-Q, and \dphase~controls rotation between Stokes-U and Stokes-V. Notice how any single parameter does not control Stokes-Q and Stokes-U. Compare with the case of circular basis where \dphase~exactly does so. Lack of the degeneracy lets us measure RM, cross hand delay at the same time. In this work for the two bursts in Band~5, we use this methodology to measure RM.

% Model
\par RM measurements are done as already mentioned above using \cmd{ultranest}. The only difference is the addition of a Delay parameter.
%Now, we attempt to model the system.
%FIXME: this "undergone" sentence structure
Starting with a pure linear polarized signal (in this formalism with Stokes-Q), first transformation undergone by the infalling radiation is the Faraday Rotation (parameterized by $\theta$). 
Then realized is the cross hand delay, which is parameterized by $\phi$.
These transformations are computed at every frequency channel with $\theta = {\rm PA} + {\rm RM} ( \lambda^2 - \lambda_c^2)$ and $\phi = f_{\rm GHz} * \pi * {\rm D}_{\rm ns}$, ${\rm D}_{\rm ns}$ denotes the cross hand delay in nanoseconds. 
\begin{equation}
    \begin{pmatrix}
        Q \\
        U \\
        V \\
    \end{pmatrix} = %
    \begin{pmatrix}
        1  &  0  &  0 \\
        0  & {\rm cos} (\phi) & -{\rm sin} (\phi) \\
        0  & {\rm sin} (\phi) & {\rm cos} (\phi) \\
    \end{pmatrix}
    \begin{pmatrix}
        {\rm cos} (2\theta) & -{\rm sin} (2\theta) & 0 \\
        {\rm sin} (2\theta) & {\rm cos} (2\theta) & 0 \\
        0 & 0 & 1
    \end{pmatrix}
    \begin{pmatrix}
        L_p \\
        0 \\
        0 \\
    \end{pmatrix}.
\end{equation}
As done with Band~4 RM measurements, we also include \cmd{equad} to expand errors since we are not actually calibrating the bursts.

\par Note that our assumption of pure Stokes-Q signal need not be the case. Our input (right-most vector) can be a mixed state but it can be always be written as a rotation matrix applied to be a pure state. 
Such a rotation matrix would be a constant and can be absorbed into the right-most matrix. Granted in such a case, recovering position angle would not be straightforward. 
But since we are not measuring position angles, we are tempted to follow this procedure for the sake of simplicity. 
% results
We apply this method to only two of the high S/N bursts. The measured RM and their uncertainties are mentioned in Sect.~\ref{ssec:mbb}. 
We make note of the large errors and attribute it to inadequate bandwidth of only $200$ MHz while measuring at high frequencies.
%The RM resolution we can expect with bursts in $1000$ and $1200$ MHz is only around 220 rad m$^{-2}$ compared with Band~4.

%\subsection{Polarization fraction measurements}
%\label{ssec:lp}

\par Prior to measuring the linear polarization fractions of the bursts, we first correct each burst with the RM measured from it using \cmd{pam} \citep{psrchive}. 
Then, we follow the procedure of \citet{EverettWeisberg2001} to de-bias and measure the linear polarization fractions of the bursts. 
This has already been specified in great detail in \citet{21NimmoR3,23BethapudiR3,24GopinathR3}, hence is omitted here.
%% or i am lazy
%% is there any point in re-writing the same shit again and again?
%% there is no point

\section{Results and discussions} \label{sec:rd}
%% how to make collage wtfffff

%\setkeys{Gin}{draft}
\begin{figure*}
    \centering
    \includegraphics[width=\textwidth,keepaspectratio]{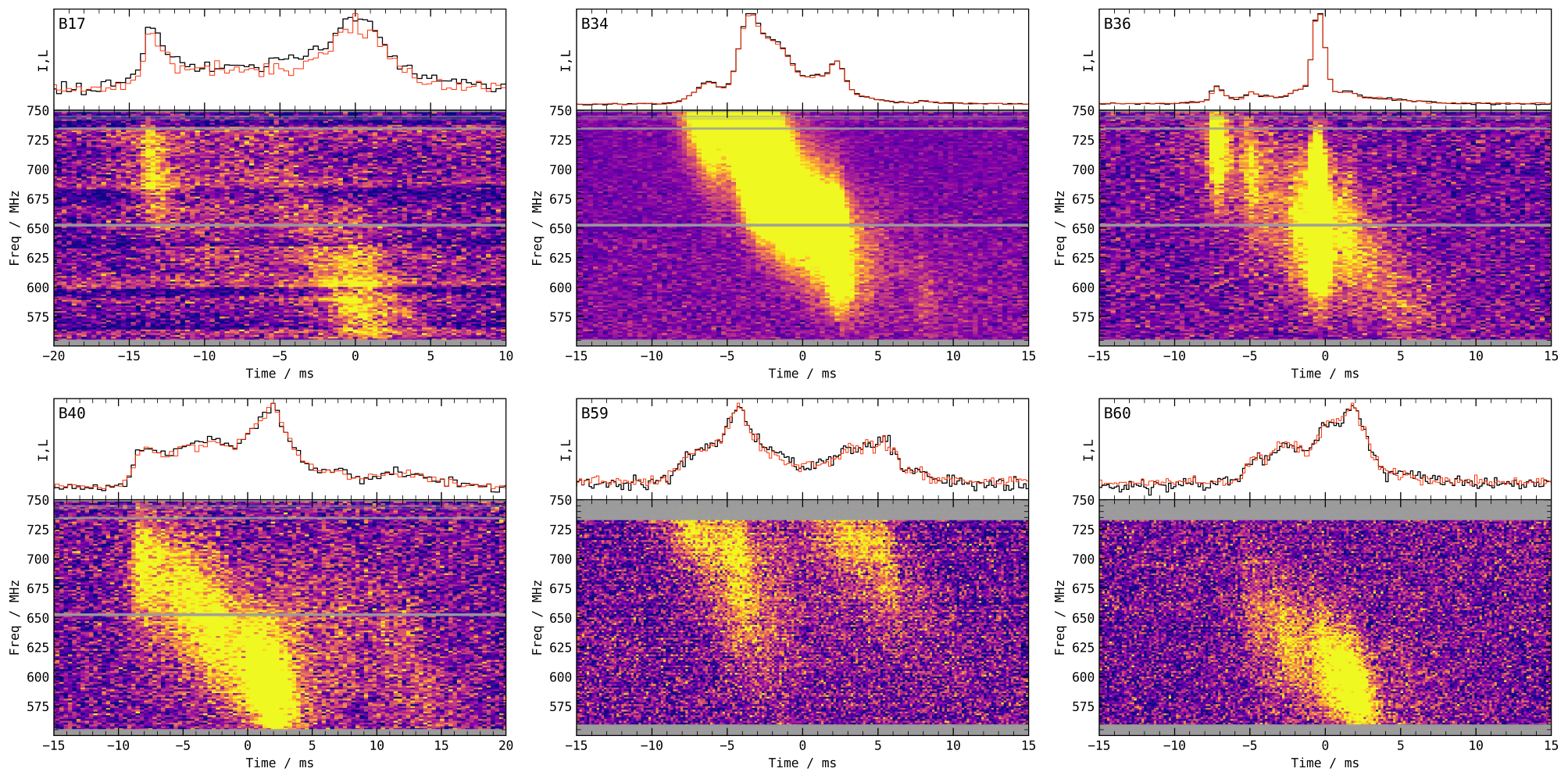} 
    \caption{A selection of polarization calibrated bursts. For each burst, we plot the de-dispersed dynamic spectra, frequency averaged Total intensity (black) and RM corrected linear polarized (red) time series. The gray bands RFI channels which have been zapped. The top left text in the time series panels is the burst id. The rest of the plots can be found in supplementary material.}
    \label{fig:bursts}
\end{figure*}
%\setkeys{Gin}{draft=false}

\subsection{Properties} \label{ssec:flurate}

\begin{figure}
    \centering
    \includegraphics[width=\columnwidth,keepaspectratio]{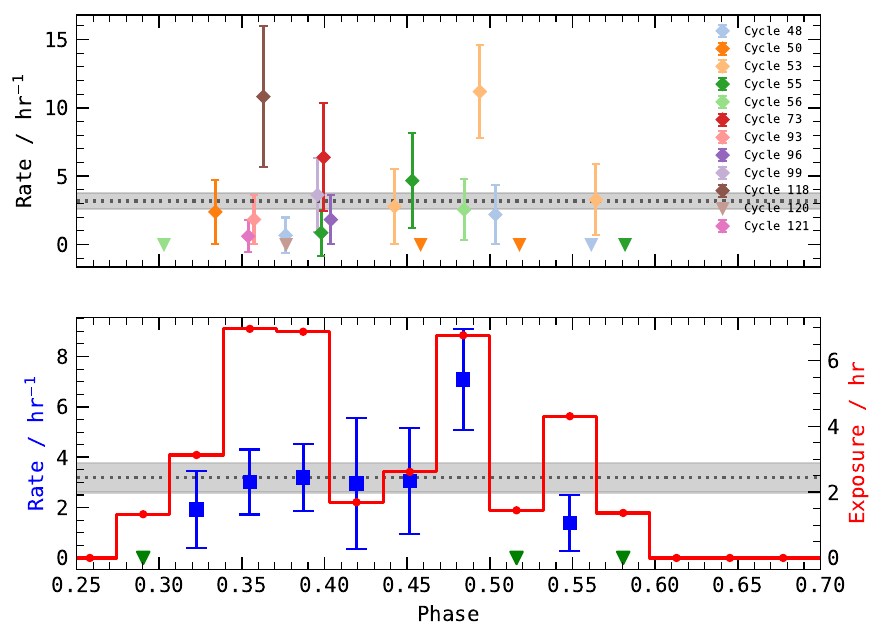}
    \caption{Rate over activity phase of \frb. 
    The horizontal dotted black line is the total average rate. The error-bars and the shaded region around the horizontal dotted line correspond to 95\% error.
    \emph{Top:} Rate per phase for each observation color-coded by Activity Cycle of the source. Detections are marked by diamonds with 95\% error bars. Non-detections are marked by triangles.
    \emph{Bottom:} Rate over activity phase when all observations are considered. See text for more details. The green triangles pointing downwards denote non-detection of bursts. The blue squares correspond to rate with 95\% error, with \emph{y}-axis towards the left. The red circles mark the total exposure at the corresponding phase and has \emph{y}-axis towards the right.
    }
    \label{fig:ratephase}
\end{figure}

%% just text for rate
%%%%% RATE
\par We report a total of $116$ bursts collected over $36.50$ hours, with which we report a total rate of $3.1 \pm 0.5$ hr$^{-1}$.
The error is $95\%$ Poissonian error.
Moreover, we compute rate from each observing session as well as the total rate and plot in Fig.~\ref{fig:ratephase} (\emph{top} panel). 
We also compute rate and total exposure as a function of Activity phase taking into account all the observations in the \emph{bottom} panel of Fig.~\ref{fig:ratephase}.
We use the latest measured period of \citet{23SandRate}~to compute phases of the bursts.
Moreover, \citet{23SandRate}~also report the period derivative which is consistent with zero, implying the computed phases of the bursts to be valid even for a large time span of data, such as our dataset.
Each observation is binned in phase-axis where every phase bin width is 0.03125 (12.25 hours). In case, any bin is partially overlapped, we add fractional weight proportional to the overlap. Thereafter, exposure and number of bursts per each phase bin is computed, which gives rate as a function of phase.
We denote the phase bins where we did not detect any bursts with green triangles.
We do not see any rate variability over phase. 
\citet{21PastorR3}~modeled rate as a function of phase using Kernel Density Estimates. 
We cannot yet perform such an analysis because of insufficient sampling of the phase region.
Quantitatively, our total observations compose of $36.5$ hours whereas the active window at $600$ MHz is about $50$ hours \citep{21ZiggyR3,23BethapudiR3}.
Therefore, we suggest we would need at least 100 hours of on source time in total before we can confidently measure phase variability \citep[][c.f.]{23SandRate}.

\begin{figure}
    \centering
    \includegraphics[width=\columnwidth,keepaspectratio]{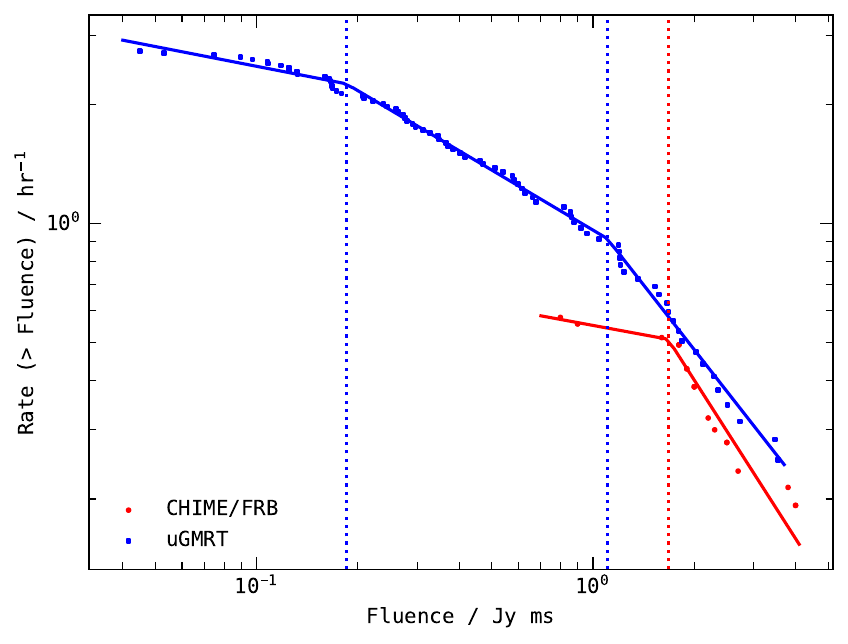}
    \caption{Rate as a function of fluence for CHIME/FRB sample \citep[][in red]{20R3Period,21ZiggyR3} and our sample (in blue). 
    Broken powerlaw are fitted to CHIME and our samples.
    The blue dotted lines are at 0.18 Jy ms and 1.10 Jy ms. 
    The red dotted line is at 1.67 Jy ms.
    See text for the $\gamma$ values.}
    \label{fig:Nflu}
\end{figure}

%%% FLUENCE 
%%%% always present our results first
\par We investigate cumulative rate distribution against fluence for our burst sample. 
Fig.~\ref{fig:Nflu} shows cumulative rate above fluence ($F$) versus fluence for GMRT-Band~4 burst sample in blue squares. 
We fit broken power-laws, where each power-law is of the form ${\rm Rate}(\geq F) \propto F^\gamma$) to it.
For our sample, we required two break points, 0.18 Jy ms and 1.10 Jy ms, to fit the data. 
Between 0.18 Jy ms and 1.10 Jy ms, we compute $\gamma=-0.51\pm0.01$, and beyond 1.10 Jy ms, $\gamma=-1.09\pm0.07$.
Of the two breakpoints, we treat the smallest fluence break point as the fluence completeness limit.
%% GMRT fluence completeness
The fluence completeness for our sample agrees well with what is computed using radiometer assuming S/N of 10, a maximum width of $40$ ms and RMS noise of $3.36$ mJy\footnote{\url{http://www.ncra.tifr.res.in:8081/~secr-ops/etc/rmsp_advanced/rmsp.html}}.
Now, we repeat the same exercise with CHIME/FRB \frb~burst sample \citep{20R3Period,21ZiggyR3}.
We plot the cumulative rate versus fluence in the same Fig.~\ref{fig:Nflu} but in red circles.
For the CHIME sample, we required only one breakpoint at 1.67 Jy ms and $\gamma$ was fitted to be $-1.34 \pm 0.11$ above it.
We note that uGMRT fluence completeness is ten times smaller than CHIME/FRB, which obviously suggests GMRT is a sensitive instrument that complements the shallow but daily observations of CHIME/FRB. 
%% PM21 also does two breaks
%Also interestingly, we note that broken power law with two breaks fits well for our sample, which is done for the first time for \frb.
%% compare with 
\citet{21PastorR3}~measures $\gamma$ to be $-1.5\pm0.2$ and $-1.4\pm0.1$ at 150 MHz (with LOFAR and above 104 Jy ms) and 1370 MHz (with APERTIF and above 7.8 Jy ms), which suggests the $\gamma$ measurements across frequencies are consistent up to $2\sigma$.
But, if it is so that $\gamma$ at 600-650 MHz is not as steep as compared to other frequencies, it could mean that GMRT is probing a deeper energy regime which is not performed by other telescopes.
%FIXME: need Laura's comment to interpret this correctly
%%

%% no DM change
%% no frequency dependency
%% LP stuff
%% depol
%% CP stuff
\par We do not observe any noticeable DM variations from the inspection of dynamic spectra of the bursts. 
This observation is strongly corroborated by \citet{23SandRate}, which also do not report any DM variation. 
Additionally, we do not observe any frequency burst envelope variations between regimes when RM was non-varying and when RM was varying \citepalias[c.f. Fig.~4]{23McKinvenR3}.
The linear polarization fractions are consistent with measurements of \citetalias{23McKinvenR3}.
Linear polarization of the bursts from this FRB undergo depolarization due to scatter in the RM \citep{22FengSRM}.
The RM scatter is only $0.12$ rad m$^{-2}$ which predicts bursts to be on average $95\%$ linear polarized at our frequency of operation (650 MHz). 
We note the inverse variance weighted mean linear polarization fraction is $95.0(1)\%$, which agrees well with the predicted depolarization.
% FIXME: LP variability is difficult to measure because 200MHz and not many bursts
Since our linear polarization fraction measurements do not hint any variability, we do not attempt to fit for RM scatter either on a per-observation basis or on the whole.
Nevertheless, we encourage the proposition of checking RM scatter variability with time, as it could provide constraint on temporal variations of RM.
%with the least polarization fraction being $55(1)\%$.
%Therefore, we postulate that the bursts in our sample which are not as linearly polarized as predicted by the model, are due to the intrinsic nature of the source.
We do not measure, and thus, report any circular polarization.

\subsection{Multi-band bursts} \label{ssec:mbb}

\begin{figure*}
    \centering
    \includegraphics[width=\textwidth,keepaspectratio]{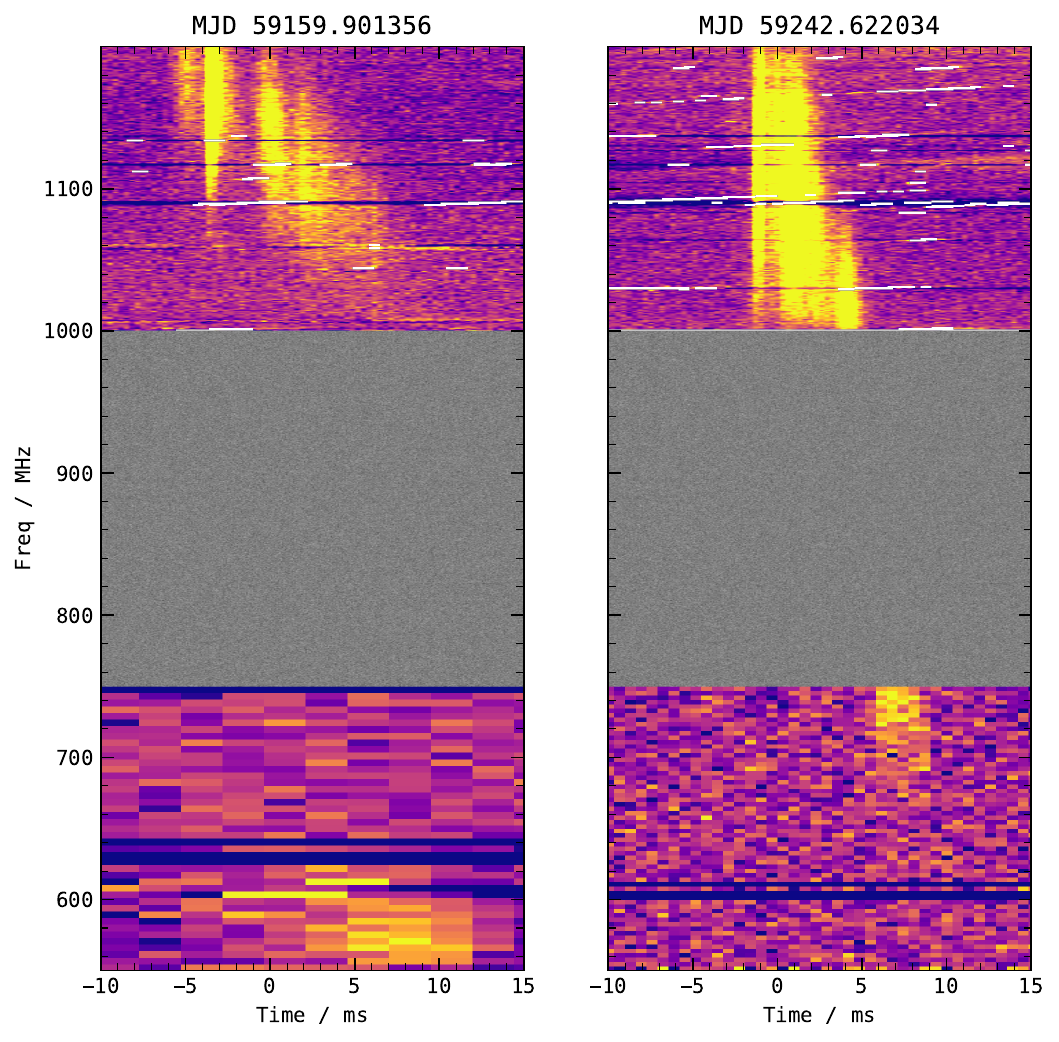}
    \caption{Time aligned filterbank data of Band~4 (550-750 MHz) and Band~5 (1000-1200 MHz) showing simultaneous detections. The filterbank data has been de-dispersed to 1200 MHz (top of Band~5). The gray region is random data plotted to show continuity. The white strips are manually flagged RFI. The time-frequency resolution and color-scale has been suitable chosen to clearly display the bursts.}
    \label{fig:multifreq}
\end{figure*}

%% intro
\par We report two instances where we detect bursts in both of the observing bands (550-750 MHz and 1000-1200 MHz) simultaneously.
We plot them in Fig.~\ref{fig:multifreq} where we de-dispersed both the bands to the same reference frequency of $1200$ MHz (top of Band~5).
In both the cases, we observe the burst is brighter in the upper band that precedes the lower band component.
In case of multi-frequency burst detected on MJD~$59242$, 
the Band~4 component to follow the structure of Band~5 component.
%\citet[Fig.~8]{23BethapudiR3}~reported a trend that bursts from \frb~have large bandwidths and smaller widths at higher frequencies compared to lower frequencies.
Moreover, this burst has a bandwidth of at least $500$ MHz and temporal width of $12$ ms.
In the multi-frequency burst detected on MJD~$59159$, the frequency structure fades towards the lower edge of Band~5 and it only reappears from the lower edge of the Band~4, implying that the frequency structure is discontinuous.
The discontinuity could be at least for $450$ MHz, as we do not know if there is a burst component in the missing frequency band from 750 MHz to 1000 MHz.
Moreover, the burst structure in Band~4 does not match well with that in Band~5.
With only two instances in the entire dataset, we refrain from discussing possible physical scenarios that could cause such bursts.
However, we detected two instances with only six hours of observing time, which suggests these instances are not a rare phenomena, therefore, we highly recommend large bandwidth observations of not just \frb~but also other rFRBs.

%% small paragraph about RM
\par We try to compare the RMs between the multi frequency components.
%We preface by drawing attention to the general agreement between CHIME/FRB bursts at 600 MHz \citepalias{23McKinvenR3} and LOFAR at 150 MHz \citepalias{24GopinathR3}.
For the multi-frequency burst on MJD~$59159$, the Band~5 RM is $-105.38 \pm 3.6$ rad m$^{-2}$.
And, for the multi-frequency burst on MJD~$59242$, the Band~5 RM is $-120.4 \pm 4.5$ rad m$^{-2}$.
We cannot measure Band~4 RM because we could not perform polarization calibration. 
For the MJD~59191 burst, we have no auxiliary scans (see Table~\ref{tab:band4}).
For the MJD~59242 burst, our \cmd{dphase} solution did not converge, and we could not generate a calibration solution.
However, we know that on these MJDs, RM had not started to change (see the following section). If we assume RM from the epoch when it was not changing \citep[$-114.6$ rad m$^{-2}$]{21NimmoR3}, the Band~5 RM measurements are within $3\sigma$.

\subsection{Band~4 RM} \label{ssec:rm}

\begin{figure*}
    \centering
    \includegraphics[width=0.95\textwidth,keepaspectratio]{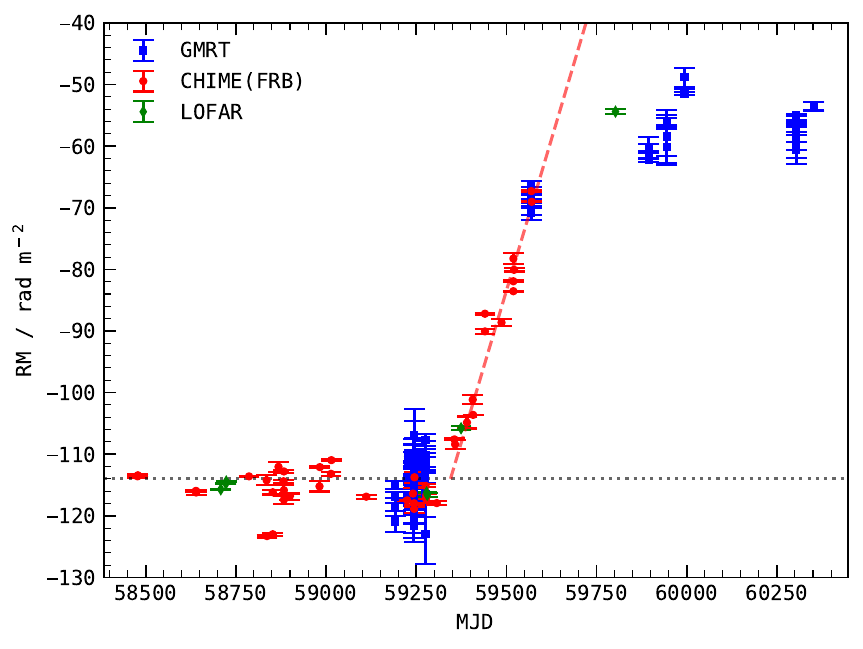}
    \caption{Observed Ionosphere corrected RM against MJD.
    CHIME/FRB \citepalias{23McKinvenR3} are in red circles. LOFAR points are in green diamonds \citepalias{24GopinathR3}. This work's points are plotted in blue squares. The dotted horizontal line is the reported RM before variability started \citep{21NimmoR3}. The dashed red line is the trend fitted by \citetalias{23McKinvenR3}.}
    \label{fig:rmmjd}
\end{figure*}

%%%%%%%%%%%%%%%%%%%%%%%%%%%%%%
%%% rewrite

% P1 - present our results, compare CHIME measurement, trend variation
\par We plot our observed ionospheric corrected RMs against MJDs in Fig.~\ref{fig:rmmjd} (in blue squares). In the same figure, we plot measurements by CHIME/FRB \citepalias[red circles]{23McKinvenR3} and LOFAR \citep[in green diamonds]{21ZiggyR3,24GopinathR3}.
We highlight the strong consistency between our RM measurements and CHIME/FRB measurements around MJD~59569.
This agreement also musters confidence in our RM measurements.
Prior to MJD~59300 the RM is not seen to vary.
From MJD~59300 until until MJD~59700, the measurements vary in a linear fashion. 
\citetalias{23McKinvenR3}~fitted a linear trend to the RM against time with slope of $0.197(6)$ rad m$^{-2}$ day$^{-1}$. 
This trend is shown with a red broken line in Fig.~\ref{fig:rmmjd}.
After MJD~59800, with one LOFAR detection \citepalias[MJD~59802]{24GopinathR3} and our detections, we see the RM measurements no longer follows the linear trend of \citetalias{23McKinvenR3}.
%The MJD~59802 LOFAR detection does not follow the linear RM trend, however, our measurements between MJD~59894 and MJD~59993 seem to have resumed a similar increasing trend.
%Our measurements from MJD~59894 and MJD~60352 suggest RM has ceased to vary.
When seen everything, we note that RM was exhibiting stochastic variability from MJD~58500 until MJD~59300, then from MJD~59300 to MJD~59700, it exhibited a secular variation, but from MJD~59800 until the end of our measurements - MJD~60352, the RM is seen to vary only stochastically.
We also note that as of now, the stochastic variability before and after the linear trend is of the same duration.
%Furthermore, at the time of writing, the RM has been varying for the \emph{almost} the same time span as it has been non-varying (for around 800 days).
Future observations would be crucial not only to observe if the increasing trend resumes but also to see if RM ever goes back to its original value.
%Should the RM go back to its original value (as during the non-varying phase), it could imply the passing of a magnetic field structure.
%And if the RM variation is periodic, it could i
%\SB{this is where it becomes really yucky-it is difficult to comment anything without being speculative. i am just commenting it out.}

% P2 - secular RM variation, compare with R1 and GCMag and sign flip
\par \citet{23McKinvenFRB}~shows that most rFRBs exhibit secular RM evolution, often with $\frac{\Delta {\rm RM}}{\rm RM} \sim 1$.
This is also seen in \frb. 
%But \citet{23McKinvenFRB}~only has few FRBs with large time span of measurements which does not help in making robust comparisons.
Futhermore, \frb~is one of the few sources which shows time-spans of secular RM variation as well as time-span of no RM variation.
%Other such sources are FRB~20121102A \citep{14SpitlerR1} and PSR~J1745-2900 \citep{13EatoughGCMag}.
Other such sources are:
PSR~J1745-2900 \citep{13EatoughGCMag} exhibits RM variability, which parallels \frb~\citep[][see Fig.~3]{18DesvignesGC}.
The RM of PSR~J1745-2900 was also seen to vary in a linear fashion \citepalias[c.f.]{23McKinvenR3}.
But, further monitoring shows that linear trend ceased but the RM is still continuing to increase (Gregory Desvignes, priv. comm.).
FRB~20121102A \citep{14SpitlerR1} has large absolute RM ($\sim 10^5$) and has exhibited decreasing RM evolution. And, it also has epochs where the evolution halts \citep{21HenningR1,22PlavinR1}.
There is also FRB~20190520B \citep{22NiuRfast}~has an extreme $|$RM$|$ value and variability. 
It has already undergone two sign flips within a span of 200 days \citep{23AnnaRfast}. 
%Such drastic variability has not been seen by any FRBs.
FRB~20180301A \citep{20Luo0301}~in addition to RM variability with one sign flip, also shows clear Dispersion Measure (DM) variability \citep{23Kumar0301}.
FRB~20201124A \citep{22LanmanR67}~showed substantial short term variability \citep{21HenningR67,22XuR67} but only for a span of forty days and became quiescent after.
%FIXME: quantify "but only for a while" 
On the contrary, FRB~20220912A does not exhibit any RM variability when observed with FAST over two months time scale \citep{23ZhangR117}.
%The RM variability of PSR~J1745-2900 parallels \frb.
%Initially, PSR~J1745-2900's RM increased slowly with time, followed by a sudden change to a more rapid, linear increase \citep[][see Fig.~3]{18DesvignesGC}.
%Further monitoring has showed that the linear trend has stopped and the RM is now increasing without a linear trend (Gregory Desvignes, priv. comm.).
%% BMAG estimate
\par We investigate if the increasing RM trend shows sign reversal.
The Milky Way contribution of RM is estimated to be $-98\pm40$ rad m$^{-2}$ \citep{22Faraday2020}.
The largest predicted Milky Way contribution is $-58$ rad m$^{-2}$ and the largest ionospheric RM corrected RM measured (on MJD~59993) is $-48.8\pm1.56$ rad m$^{-2}$.
There is a tentative detection of sign flip.
%, we can report a sign flip in the host galaxy RM with a significance of $1.2\sigma$. 
%We note that significance of the sign flip can be enhanced if future RM measurements are even larger or if Milky Way RM model become more precise.
%The former requires future measurements and the latter is simply beyond the scope of this paper.
% here SIGNFLIP
There are three possible scenarios for observed RM trend - (i) changing electron density, (ii) changing path length of the magnetized Faraday screen, and (iii) changing magnetic field \citep[also see][]{23McKinvenFRB}. 
Changing electron density or path lengths of the active media along the line-of-sight would cause correlated variations in the DM, which neither \citetalias{23McKinvenR3} or \citet{23SandRate} or we observe.
%The Faraday width (or $\sigma_{\rm RM}$) along the line-of-sight is only $0.12$ rad m$^{-2}$ \citep{22FengSRM}, which is only two percent of the latest measured RM.
%Such a low Faraday width suggests the line-of-sight to have only a thin-Faraday screen which rules out changing path length to cause RM variations. \SB{confirm this with Ann Mao}
%We also suggest measuring the variability in the $\sigma_{\rm RM}$ to see if it is co-variant with RM. Such a covariance would strongly support changing Faraday screen to cause varying RM measurements. But these measurements could only be robustly computed at lower frequencies since the gradient of depolarization (as a function of frequency) is maximum there.
The presence of a sign flip in the RM would vastly suggest that variations seen are due to changing magnetic field.
Future observations will be extremely crucial in probing if the tentative detection of sign flip is, infact, genuine.
If that is the case, it would make \frb~the third FRB, after FRB~20190520B \citep{23AnnaRfast} and FRB~20180301A \citep{23Kumar0301}, to show a sign reversal.
Lastly, we strongly urge effort to study the large scale magnetic field structure of the FRB host galaxy. 
Such a study would provide an alternative line of evidence for the average magnetic field measurements, and would elucidate the role past mergers \citep{22KaurR3} may have played in the creation of FRB source.
%We plot the RM$_{\rm host}$ in the \emph{right} y-axis of Fig.~\ref{fig:rmmjd}. For that, we used $RM_{\rm MW} = -98$ rad m$^{-2}$.

% P3 - PRS, models
%% PRS models are 
%% R1 RM was varying -> R1 has PRS -> PRS models appear
%% R3 appears -> R3 starts varying its RM -> we cannot say no PRS.
\par There are multiple models involving Persistent Radio Sources (PRSs) at the FRB location which predict RM (and DM) evolution.
Such models have been proposed for FRB~20121102A, FRB~20190520B, and FRB~20201124A, all of which have PRS detections \citep{chatterjee2017,22NiuRfast,23BruniPRS}.
For instance, \citet{18PiroGaensler}~postulates an expanding shocked supernovae remnant (SNR) to cause DM and RM evolution.
The FRB emitting object is within the SNR and the shocked boundary between the ejecta and background ISM provides necessary conditions to affect DM and RM with time.
Similarly, \citet{Margalit2018}~propose the FRB source, a magnetar, is embedded in a magnetized wind nebula (MWNe), which causes the DM and RM variations.
Also, interestingly, \citet{23BruniPRS}~which also attributes RM evolution with PRS, proposes a correlation between luminosity of the PRS and the magnitude of the observed RM.
%\citet{23BruniPRS}~suggest the magnitude of RM is related with the luminosity of the PRS.
This relation holds for FRB~20121102A and FRB~20190520B, both of which have $|$RM$|$ in $10^5$ rad m$^{-2}$ \citep{21HenningR1,23AnnaRfast}.
And also for FRB~20201124A whose $|$RM$|$ is order of 650 rad m$^{-2}$ \citep{21HenningR67,22XuR67}.
%\citet{23BruniPRS}~also report PRS detection for FRB~20201124A 
However, there has only been an upper limit on the luminosity of \frb~\citet{20MarcoteR3}, which implies the magnitude of RM of \frb~is low. This adds weight to the model.
%The applicability of such models to \frb~is doubtful.
A common feature for the above models is the predicted monotonic RM trend. 
Testing for monotonic nature of the RM variation can be easily performed by future measurements.
But this could only form a weak evidence for presence of PRS.
A lack of robust detection of PRS, as in case of \frb, makes applying RM evolution models, which require a PRS, a bit far fetched.
\par Lastly, we highlight that on \texttt{29jan2021}, we observed a rate of $11.4 \pm 3.4$ hr$^{-1}$.
The \texttt{29jan2021} rate is almost $5\sigma$ larger than global rate.
%%Although we note that within 14 observations, observing such a high rate in one observation is still statistically consistent.
Thereafter, the RM variability was seen. 
We explore the possibility that this could be more than coincidence.
On one hand, the rate increase is statistically significant.
But, the \texttt{29jan2021} observation has around four hours of on source time which is twice the usual on source time. 
A period of non variability and a period of variability which are demarcated by a short period of heightened activity sounds plausible that heightened activity could also have been caused by the same mechanism which is causing the observed RM trend.
At this point, we merely make this speculation and hope future observations when such similar heightened rates are visible would make this connection more evident.

\section{Conclusions} \label{sec:con}

% just start with itemize
%We present the following results in this paper:
\par Firstly, we devise multiple strategies to polarization calibrate the bursts in the dataset.
Each strategy is designed to make use of the available auxiliary scans taken during the observing session (Sect.~\ref{ssec:polcal}).
Then, we implement a pipeline to measure RM of the polarization calibrated bursts using $QU-$fitting (Sect.~\ref{ssec:rmm}). 
Using which, we first correct for the Faraday Rotation and measure polarization properties of the bursts (Sect.~\ref{ssec:rmm}).
We also verify the calibration solution using pulsar scans, and comment on how calibration solution would affect subsequent RM measurements (Sects.~\ref{a:sec:rmver} and \ref{a:sec:dpvar}).

\par From our observing campaign, we detect 116 bursts with 36.5 hours on source spanning over 1200 days.
%Thereafter, we present polarimetric results.
Following are our findings:

\begin{itemize}

%% MULTIBURST
\item For few of the sessions, we performed simultaneous observations of \frb~in Band~4 (650 MHz) and Band~5 (1.1 GHz). 
We leave the proper calibration of Band~5 bursts to a future work but we have tried to measure RMs of the two uncalibrated brightest bursts in our sample (Sect.~\ref{ssec:rmm}).
Bursts show varied structures and fact that we detected two bursts within six hours of observing advocates future large bandwidth observations (Sect.~\ref{ssec:mbb}).

%% RM(TIME)
\item We plot the time variability of RM in Fig.~\ref{fig:rmmjd}. The linear trend seen by \citetalias{23McKinvenR3} is no longer consistent with the data, and the source has entered a regime where the RM is only varying stochastically.
%We estimate the parallel average magnetic field along the line-of-sight and comment that it has undergone at least 350\% fractional change over a span of 800 days.
We report a tentative detection of sign flip in RM strongly suggesting the RM variations to be driven by magnetic field variations.
Future observations will trace the RM variations and would certainly provide new insights into possible scenarios.
%Continuing increasing trend would help place constraints on SNR and MWNe models.
%However, we are severely limited in our constraints due to non-detection of PRS.
Lastly, we motivate large scale magnetic field study of the FRB host galaxy to elucidate any preferred magnetic field structure that could be causing the observed RM variations.
\end{itemize}

\section{Acknowledgements}

SB would like to thank a lot of people: Simon Johnston for useful discussions about polarization calibration strategy, Frank Schinzel for helping understand time variability of calibration solutions, Gregory Desvignes for extensive and exhaustive help with the manuscript.
We thank the staff of the GMRT that made these observations possible. 
VRM gratefully acknowledges the Department of Atomic Energy, Government of India, for its assistance under project No. 12-R\&D-TFR-5.02-0700.
GMRT is run by the National Centre for Radio Astrophysics of the Tata Institute of Fundamental Research.
LGS is a Lise Meitner Independent Max Planck research group leader and acknowledges support from the Max Planck Society.
Part of this research was carried out at the Jet Propulsion Laboratory, California Institute of Technology, under a contract with the National Aeronautics and Space Administration.

\section{CODE and DATA AVAILABILITY}
All the code for GMRT Coherence data reduction and analysis, and to reproduce the results and plots and tables are given in the following hyperlinks - \href{https://github.com/dongzili/GMRT-FRB}{GMRT-FRB} and \href{to-be-made-public}{GMRT-R3-RM} respectively. 

\bibliographystyle{aa}
%\bibliography{frb20201124a}
\bibliography{cal,rest,frb}

\appendix

\section{Verification of calibration} \label{apex:vcal}

\subsection{RM verification} \label{a:sec:rmver}

\begin{figure}
    \centering
    \includegraphics[width=0.9\columnwidth,keepaspectratio]{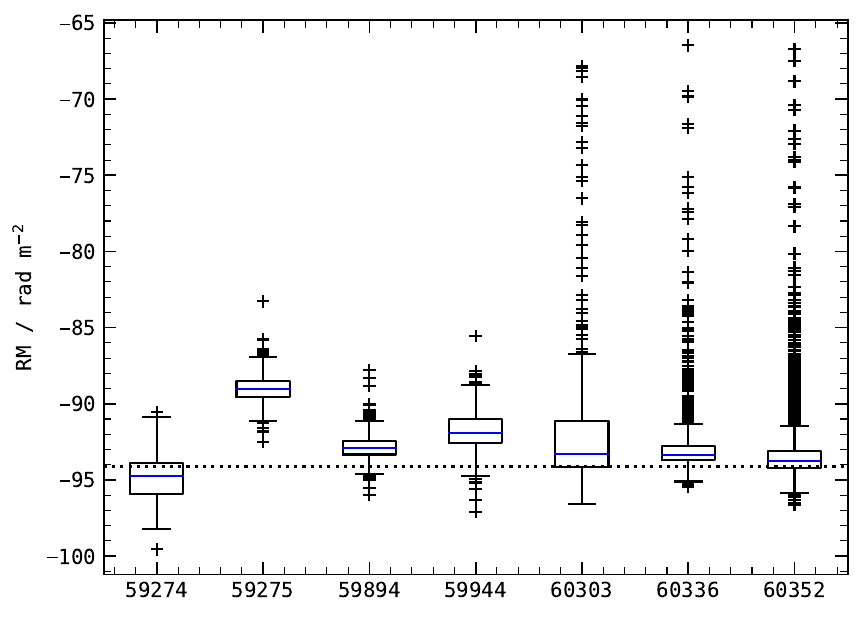}
    \caption{Boxplot of ionospheric corrected RM measurements of single-pulses from PSR~J0139+5814 taken on different days. The blue line in each box denotes the median. Each box spans for Inter-Quartile-Range. The horizontal dotted line is the literature RM value for the pulsar. The large number of outliers on MJDs 60303, 60336 and 60352 is due to the heightened ionospheric activity reported in the GMRT Cycle~45 observations.}
    \label{fig:rmboot}
\end{figure}

\par We have seven observations of PSR~J0139+5814. We perform a single-pulse search (as described in Sect.~\ref{sec:obssearch}), polarization calibrate (see Sect.~\ref{ssec:polcal}), and measure RM (Sect.~\ref{ssec:rmm}) from each of the single pulses.
We use this dataset of single-pulses to verify RM measurements.

\par We plot boxplot of the measured ionospheric corrected RM in Fig.~\ref{fig:rmboot}. We also plot the literature RM value with the horizontal dotted line.
With this plot, we want to highlight the spread in RM values seen with single pulses in a given observation.
And, the quirk that RM measurements are not completely consistent from observation to observation suggesting some systematic at play.
Similar deviation from literature value was detected for the same pulsar in \citet{GMRTPolChandra23}.
The large tail we observe in case of observations of MJD~60303, MJD~60336 and MJD~60352 is due to the strong ionospheric activity recorded in GMRT Cycle~45.

\section{\dphase~variability} \label{a:sec:dpvar}
% maybe i move this to appendix

\par RM measurements are extremely sensitive to the cross-hand delay because the degeneracy described earlier. 
Therefore, we have to exercise extreme caution in measuring the delay and uncovering any systematics.
GMRT being an interferometer where phasing operation is performed before every science scan, the \dphase~is susceptible to change on macroscopic variables (like ambient temperature and such). 
Investigating the underlying factors requires much more carefully planned study and is beyond the scope of this paper.

\begin{figure}
    \centering
    \includegraphics[width=0.85\columnwidth,keepaspectratio]{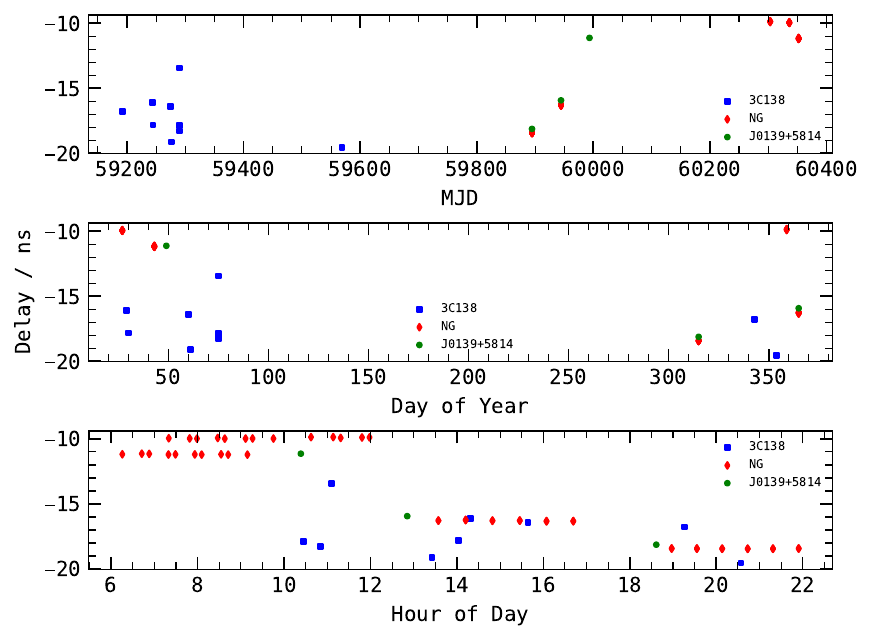}
    \caption{Cross hand delay measurements from different scans - 3C138, noise diode and PSR~J0139+5814 plotted against MJD, day-of-year and hour-of-day. The sign of the delay is purely a mathematical artefact.}
    \label{fig:delay}
\end{figure}

\par We plot the cross-hand delay we measure from scans of 3C138, noise diode and pulsar J0139+5814 against MJD, day-of-year and hour-of-day in Fig.~\ref{fig:delay}. 
The delay seems to vary from observation to observation. 
In almost all the cases, we only have a single measurement in an observation.
However, for \texttt{12nov2022} and \texttt{31dec2022}, we have noise diode scans and PSR~J0139+5814 pulsar scans.
The cross-hand delay measurement from all of them is consistent within an observation.
The standard deviation of measurements from \texttt{12nov2022} and \texttt{31dec2022} is less than $0.1$ ns.
This can also be seen in the plot in the bottom panel where points follow a horizontal line.

\par
\begin{figure}
    \centering
    \includegraphics[width=0.85\columnwidth,keepaspectratio]{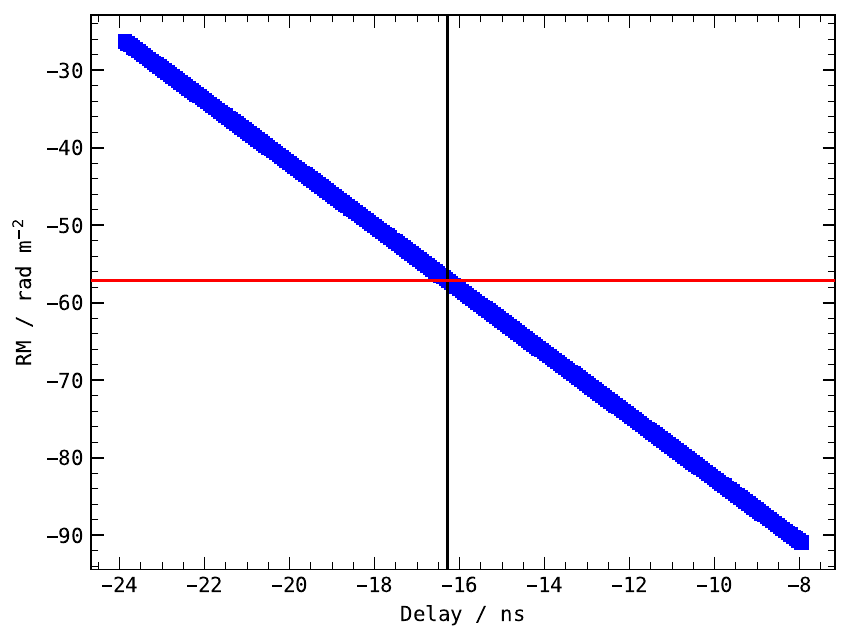}
    \caption{RM measurements when calibrating for different cross hand delay. The black vertical line is the delay measurement from noise diode scan. The red horizontal line is the RM measurement. See text for details.}
    \label{fig:rmdphase}
\end{figure}

 The fact that multiple delay measurements taken within an observation are consistent even with multiple phasings lends confidence to our RM measurements throughout the dataset.
Alternatively, we investigate how much an error in cross-hand delay affects RM measurements.
We take the \texttt{B78} burst which was observed on MJD~59944. We construct multiple calibration solutions (\cmd{pacv}) files where \g~and \dg~are the same but only the cross-hand delay (${\rm D}_{\rm ns}$) is varied. 
We apply each calibration solution to the burst and perform RM measurement as mentioned in Sect.~\ref{ssec:rmm}.
We plot the cross-hand delay versus the measured RM in Fig.~\ref{fig:rmdphase}.
In the same figure, we plot the measured RM of \texttt{B78} as red horizontal line and the black vertical line is the cross-hand delay used to calibrate this burst. 
We note that a $2$ ns difference in the delay measurement causes $\mathcal{O}(10)$ difference in RM measurement.
Nevertheless, as mentioned above, the variability we have seen in the delay measurement, which is less than $0.1$ ns, does not affect our confidence in our measurements.
%%%

%%\bsp	% typesetting comment
\label{lastpage}

\end{document}